\documentclass[journal,twoside,web]{ieeecolor}
\usepackage{tmi}
\usepackage{cite}
\usepackage{amsmath,amssymb,amsfonts}
\usepackage{algorithmic}
\usepackage{graphicx}
\usepackage{bm}
\usepackage{float}
\usepackage{breqn}
\usepackage{subcaption}
\usepackage{bm}
\usepackage{hyperref}
\hypersetup{
    colorlinks=true,
    linkcolor=blue,
    filecolor=magenta,      
    urlcolor=cyan,
    pdftitle={Overleaf Example},
    pdfpagemode=FullScreen,
    }
\usepackage[linesnumbered,ruled]{algorithm2e}

\usepackage{soul}

\usepackage{textcomp}
\def\BibTeX{{\rm B\kern-.05em{\sc i\kern-.025em b}\kern-.08em
    T\kern-.1667em\lower.7ex\hbox{E}\kern-.125emX}}
\begin{document}
\title{Neural Computed Tomography}
\author{
Kunal Gupta, Brendan Colvert, and Francisco Contijoch
\thanks{This work was supported by NIH grant HL143113.}
\thanks{Kunal Gupta (Dept. of Computer Science Engineering), Brendan Colvert (Dept. of Bioengineering), and Francisco Contijoch (Dept. of Bioengineering, Radiology) were with the University of California San Diego. B.C is now with Medtronic. (e-mails:k5gupta, bcovlert, and fcontijoch@ucsd.edu).}
}

\maketitle

\begin{abstract}
Motion during acquisition of a set of projections can lead to significant motion artifacts in computed tomography reconstructions despite fast acquisition of individual views. 
In cases such as  cardiac imaging, motion may be unavoidable and evaluating motion may be of clinical interest. 
Reconstructing images with reduced motion artifacts has  typically been achieved by developing systems with faster gantry rotation or using algorithms which measure and/or estimate the displacements.
However, these approaches have had limited success due to both physical constraints as well as the challenge of estimating/measuring non-rigid, temporally varying, and patient-specific motions. 
We propose a novel reconstruction framework, NeuralCT, to generate time-resolved images free from motion artifacts. 
Our approaches utilizes a neural implicit approach and does not require estimation or modeling of the underlying motion. 
Instead, boundaries are represented using a signed distance metric and neural implicit framework.
We utilize `analysis-by-synthesis' to identify a solution consistent with the acquired sinogram as well as spatial and temporal consistency constraints. 
We illustrate the utility of NeuralCT in three  progressively more complex scenarios: translation of a small circle, heartbeat-like change in an ellipse's diameter, and complex topological deformation. 
Without hyperparameter tuning or change to the architecture, NeuralCT provides high quality image reconstruction for all three motions, as compared to filtered backprojection using  mean-square-error and Dice metrics. Code and supplemental movies are available at \href{https://kunalmgupta.github.io/projects/NeuralCT.html}{https://kunalmgupta.github.io/projects/NeuralCT.html}.

% Code and data will be publicly released upon acceptance. 
\end{abstract}
% \vspace{-3pt}

\vspace{-15pt}
\section{Introduction}
\label{sec:introduction}
X-ray computed tomography (CT)'s ability to noninvasively and volumetrically evaluate patient anatomy with high spatial resolution has led to a widespread clinical use.
Unfortunately, image quality can be reduced if object motion occurs during the acquisition.
This is particularly concerning when imaging fast moving structures (e.g., the heart) as current single-source conebeam CT scanners require 200-300 ms to perform a full gantry rotation.
As a result, while CT is used for the evaluation of suspected coronary artery disease \cite{neglia2015detection} and acute chest pain \cite{hoffmann2012coronary}, even small motions ($\sim$1.2mm) lead to significant blurring of key structures such as coronary arteries \cite{contijoch_medphys_2017}.
Further, motion of highly-attenuating, metallic devices can further impair clinical assessment \cite{de1998metal}.
In addition to static, single frame imaging, there is growing interest in using CT to evaluate heart dynamics \cite{maffei2012left,pourmorteza2012new,pourmorteza2016correlation}.
Various techniques have been developed to avoid, reduce, and/or correct motion artifacts.
Broadly, these approaches aim to either reduce the amount of data needed for image reconstruction or correct for the underlying object motion \cite{Flohr2019}.
However, as described in Section \ref{sec:related}, current methods are limited.

In this work, we propose a new approach for time-resolved reconstruction of CT images of moving objects without estimation or correction of object motion.
We do so by developing a novel neural network architecture.
Neural networks have proven useful in computer vision and medical imaging learning problems such as classification \cite{class1,class4} and segmentation \cite{seg1,seg3}. 
In addition, neural networks can serve an efficient framework for representation of complex scenes (i.e., neural representation) \cite{mildenhall2020nerf}. 
A key benefit of neural representations is that they are highly flexible, can represent arbitrarily complex scenes, and can be easily optimized via gradient descent.  

Recently,  neural representation has been combined with volumetric differentiable renderers \cite{mildenhall2020nerf,wang2021neus} to perform optimization directly from 2D images (projections). The CT image reconstruction problem is analogous to accurately solving the inverse volumetric object rendering problem dealt in previous works. 

Therefore, in this work we test the \textit{hypothesis} that a neural representation-based image reconstruction algorithm is capable of producing time-resolved images from conventional CT data (sinograms) and when applied to imaging data of moving objects, can reduce motion artifacts as compared with conventional reconstruction without the need for an explicit motion model.

In our framework, we utilize a boundary-based representation and jointly estimate the position and displacement of object boundaries to facilitate optimization.
Specifically, we represent our target image as objects-of-interest with temporally-evolving boundaries instead of as scalar-valued intensity image or set of images. 
We do so by using the signed distance function (SDFs) as a spatiotemporal implicit representation. 
Our optimization then solves for the implicit representation which is most consistent with the acquired sinogram.
Instead of using data-driven priors, we promote convergence towards low frequency solutions which results in physically plausible reconstructions. 
This is achieved by using Fourier coefficients to represent the temporal evolution of the SDF. 
Therefore, our approach does not assume any type of object motion nor require additional data (additional CT projections, motion field estimates, or physiologic signals such as the ECG).
We find our approach provides a flexible method to accurately resolves motion of object boundaries.

Our contributions are as follows:
\begin{itemize}
    \item We pose CT image reconstruction as an object boundary-based problem and show that this enables accurate reconstruction of moving objects.
    \item We present a pipeline, NeuralCT, which combines neural rendering with implicit representations to reconstruct objects undergoing fast and complex deformations. 
    
    % solve the challenging inverse problem — reconstructing images of moving objects from CT data.
    \item We demonstrate that NeuralCT can reconstruct movies of moving objects imaged with standard axial CT acquisitions without assuming priors related to the motion. 
    % \item We show that NeuralCT can accurately resolve fast and complex deformations.
    \item We highlight that NeuralCT is robust to the choice of hyperparameters and can be readily used to solve for a wide range of motions 
    % without the need for hyperparamter optimization.
    % \item We have made the implementation of NeuralCT available at: (github repository made available upon acceptance).
\end{itemize}

This paper is organized as follows. Section II discusses prior works related to NeuralCT. A component-wise description of NeuralCT is covered in Section III followed by implementation details in Section IV. Sections V and VI respectively detail the experiments and results. 

% \vspace{-5pt}
\section{Related Works}
\label{sec:related}
\subsection{Avoidance of Motion Artifacts}
Image artifacts can be minimized by limiting motion during the acquisition.
When possible, anatomy of interest can be held still. 
When imaging the heart, cardiac motion can be reduced by imaging during portions of the cardiac cycle with relatively small motion (either the end-systole or diastasis periods) \cite{lin2009basic}.
However, this can be limited in noncompliant patients or in the case of heart imaging, in patients with high heart rates.
Further, static imaging precludes assessment of important motion such as joint articulation or cardiac dynamics.

Faster gantry rotation speeds can also be used to reduce motion artifacts. 
Increased rotation speed for single-source systems is mechanically challenging and requires an increase in x-ray tube power to maintain similar image quality \cite{Flohr2019}.
However, dual-source systems have been shown to improve the temporal resolution \cite{flohr2006first,mccollough2008measurement}.
Unfortunately, the increased mechanical complexity or additional sources has limited design of systems with more than two sources \cite{besson2015new}. 
Nonmechanical, electron beam CT systems, initially introduced in the 1980s, can achieve fast acquisition times ($\sim$50ms) but suboptimal image quality has limited routine clinical use \cite{agatston1990quantification,budoff1996ultrafast}.
As a result, current single-source conebeam CT systems reconstruct images with temporal footprints $>$50ms which can result in  artifacts due to heart motion. 

\vspace{-10pt}
\subsection{Motion Estimation Approaches}
Correcting motion artifacts by jointly estimating image intensity and motion over time on a per-pixel basis is significantly underdetermined. 
Therefore, estimating motion using computer vision approaches can prove beneficial (e.g., \cite{isola2008motion,tang2012fully} ).
For example, partial angle images can be used to estimate and correct for temporally-constant but spatially-varying motion \cite{hahn2016reduction}.
One such approach (``SnapShot Freeze", GE Healthcare) has been developed into a clinical solution and been shown to improve imaqe quality \cite{sheta2016impact} and reduce the presence of artifacts \cite{sheta2017impact}.
However, validation of these methods has been limited especially when correcting complex motions.
However, the development of advanced phantoms such as XCAT \cite{segars20104d} have recently enabled improved evaluation \cite{lee2020validation}.
In addition, recent work has leverage machine learning to improve motion estimation \cite{maier2021deep}. 

\vspace{-10pt}
\subsection{Motion Artifact Removal}
Machine learning has also been used to correct motion artifacts in reconstructed images.
For example, Ko et al developed a deep convolutional neural network (CNN) to compensate for both rigid and nonrigid motion artifacts \cite{ko2021rigid}.
Similarly, Lossau et al used three CNNs to reduce the metal artifacts created by pacemakers \cite{lossau2020learning}.
The classification of coronary artery plaques blurred by motion has been improved by fine tuning of inception v3 CNN \cite{zhang2020motion}.

\vspace{-8pt}
\subsection{Neural Implicit Representations and Reconstruction}
There has been a large body of work exploring neural implicit representations (NIRs) for efficient storage of high resolution information. 
This is due to the fact that NIRs are continuously differentiable, which allows for theoretically infinite resolution and efficient optimization using classic gradient descent techniques \cite{kingma2014adam}. 
In contrast, conventional representations require significantly higher memory (voxels \cite{riegler2017octnet,gupta2021octree}), make compromises on topology information (point clouds \cite{qi2017pointnet}) or give unrealistic surfaces post optimization (meshes \cite{gupta2020neural}).

% There has been a large body of work exploring alternative neural implicit representations (NIRs) for conventional scene representations such as point clouds, voxel grids, and meshes.

% This is due to the fact that NIRs are continuously differentiable, which allows for efficient optimization using classic gradient descent techniques \cite{kingma2014adam}. 

% NIRs can also efficiently store high resolution information, relative 

% to conventional representations which require significantly higher memory (voxels \cite{riegler2017octnet,gupta2021octree}), make compromises on topology information (point clouds \cite{qi2017pointnet}) or give unrealistic surfaces post optimization (meshes \cite{gupta2020neural}).

Since classic shape reconstruction works by Park et al. and Mescheder et al. \cite{park2019deepsdf,mescheder2019occupancy}, NIRs have been used for novel view synthesis \cite{sitzmann2019scene,liu2020neural,lombardi2019neural,mildenhall2020nerf} and multi-view reconstruction \cite{jiang2020sdfdiff, saito2019pifu, saito2020pifuhd}. 
Recently, NIRs have been used to improve CT \cite{sun2021coil,wu2021arbitrary,tancik2021learned,lindell2021autoint,shen2021nerp} and MR\cite{wu2021irem} imaging.
The use of NIRs to help solve inverse problem of CT reconstruction was first shown by \cite{sun2021coil}. 
Such techniques have since shown to enable super resolution \cite{wu2021arbitrary} and improve sparse CT reconstruction \cite{tancik2021learned,lindell2021autoint,reed2021dynamic,shen2021nerp}. 
Broadly, these approaches incorporate the data acquisition process to optimize the NIRs for static objects. 
% The methods in these works enable static object reconstruction. 
In contrast, the focus of this paper is to reconstruct objects that move during data acquisition. 
While prior methods have extended NIRs to dynamic scenes \cite{reed2021dynamic, park2020deformable, xian2021space, li2021neural, Pumarola_2021_CVPR}, they seek to model the temporal change in intensity that is a step function for a  moving object with uniform intensity. 
Our key insight is to instead represent object boundaries as SDFs, a physically motivated representation, to accurately capture boundary motion and enable time-resolved reconstruction. 
Moreover, we demonstrate that our technique can utilize the result of static reconstruction methods (in our case, filtered backprojection) as an initialization to reduce motion artifacts.  

% do not exploit the fact that moving objects of uniform intensity can be represented as boundaries whose location changes smoothly over time instead of voxels with temporally-varying intensities. 
% Our method models objects as SDFs, a physically motivated representation, to accurately capture boundary motion and enable time-resolved reconstruction. 
% Moreover, we demonstrate that our technique can be applied to the result of static reconstruction methods (in our case, filtered backprojection) to reduce motion artifacts.  

% \vspace{-5pt}
\section{Formulation of NeuralCT}
\label{sec:formulation}
Here, we describe two key components of our approach, \emph{implicit representation} and \emph{differentiable rendering}, before outlining the specific design of NeuralCT in Section IV.

\vspace{-10pt}
\subsection{Implicit Representations and Inverse Problems}
% Representation Overview
3D medical images are conventionally represented using volume elements (voxels), which are a direct extension of 2D picture elements (pixels).
However, voxel-based represention of high resolution objects is computationally expensive\cite{riegler2017octnet,gupta2021octree}. 
Meshes are an alternative which can represent the `surface' information of a 3D object using a piece-wise triangulation. 
However, once a mesh is defined, topological issues can arise upon deformation\cite{gupta2020neural}. 

%Implicit representation
As their name suggests, implicit representations seek to store information implicitly as a zero level set of a higher dimensional function. 
This can overcome the limitations associated with both voxel- and mesh-based representations. 
Specifically, the signed distance of a point in 3D space to a nearby surface can be used as a representation which avoids the topological issues associated with meshes as the signed distance map can be deformed without creating ``holes" or ``breaks" in the surface.
Further, encoding the signed distance map with neural network enables a compressed representation of the object\cite{davies2020effectiveness}.

In this work, we implicitly represent a scene using a vector of signed distance functions (SDF) $\mathbf{f}:\mathbb{R}^N \times \mathbb{R} \rightarrow \mathbb{R}^{K}$ to map the spatiotemporal coordinate $\langle \mathbf{x},t \rangle \in \mathbb{R}^N \times \mathbb{R}$ to the boundaries of $K$ objects in spacetime represented by their SDF values $\mathbf{f}(\mathbf{x},t) \in \mathbb{R}^K$ where $N$ is the number of spatial dimensions and $K$ is the number of objects represented. 
The key benefit of using an SDF representation is that non-boundary locations provide information about the boundary displacement.
This avoids having to explicitly track the boundary displacement, as is the case with other representations such as meshes or occupancy functions.
 
%Neural Implicit
Numerically solving an implicit representation inverse problem requires a discrete representation.
One conventional approach is to transform the object representation into a dense set of scalar values (e.g., a voxel grid).
However, fitting a neural network to the representation is an efficient and differentiable alternative. 
Further, a neural network compactly adapts the spatiotemporal resolution of the representation to the complexity of the underlying object.
In this work, we approximate the SDF representation $\mathbf{f}(\mathbf{x},t)$ by a neural network $\hat{\mathbf{g}}(\mathbf{x},t;\mathbf{w})$, where $\mathbf{w}$ is a set of network parameters.

\vspace{-10pt}
\subsection{Optimization via Differentiable Rendering}
In the absence of a known ground truth, `analysis-by-synthesis' methods can be used to identify a solution that explains the observed data (in a different domain) \cite{analysisbysynthesis}.
Differentiable rendering is one such technique and is conventionally used to identify the shape $\hat{S}$ of an object that best `explains' its acquired projection set $\{P_1, P_2, \cdots P_N\}$.
For a shape $\mathcal{S}$, we achieve this by minimizing the error defined as the norm of differences between predicted and observed projections, i.e. 
\begin{equation}
    \hat{S} = \underset{s}{\arg\min} \sum_{i=0}^{N}||P_i - R(s;\theta_i)||_p
\end{equation}

Here $R(\mathcal{S};\theta_i)$ is the rendering operator that projects a shape $\mathcal{S}$ from `shape space' to `projection space' based on CT gantry position $\theta_i$. 
To enable gradient backpropagation and optimizing via standard gradient descent methods, the rendering operator $R$ should be differentiable \cite{kato2020differentiable}.

% \vspace{-5pt}
\section{NeuralCT}
\label{sec:NeuralCT}
Here, we describe the core components of NeuralCT.

\vspace{-10pt}
\subsection{Architecture}
As noted in Section \ref{sec:formulation}, we approximate the spatiotemporal SDF function $\mathbf{f}(\mathbf{x},t)$ of an object via a neural network $\mathbf{\hat{g}}(\mathbf{x},t;\mathbf{w})$.
We do this by creating two sub-networks that jointly approximate the stationary SDF $\mathbf{\hat{g}_{E}}$ as well as its spatiotemporal evolution $\mathbf{\hat{g}_{V}}$. 

\textbf{EncoderNet}
For each location $\mathbf{x} \in \mathbb{R}^N$, the stationary component of the object's SDF was represented using a network EncoderNet $\mathbf{\hat{g}_E}(\mathbf{x};\mathbf{w_E}):\mathbb{R}^N \rightarrow \mathbb{R}^K$. 

\textbf{VelocityNet} 
For each location $\mathbf{x} \in \mathbb{R}^N$, the temporal evolution of an object's SDF was parameterized using Fourier coefficients $ \{\mathcal{A}_0, \mathcal{A}_1, \cdots, \mathcal{A}_{MK}, \mathcal{B}_0, \mathcal{B}_1, \cdots, \mathcal{B}_{MK}\}$ where $(\mathcal{A}_i, \mathcal{B}_i)$ respectively are the coefficients for sines and cosines.
The coefficients were fit by a network $\mathbf{\hat{g}_F}(\mathbf{x},t;\mathbf{w_F}):\mathbb{R}^N \rightarrow \mathbb{R}^{2MK}$.
Then, $\mathbf{\hat{g}_V}(\mathbf{x},t; \mathbf{w_F})$ was be computed as:

\begin{dmath}
    \mathbf{\hat{g}_{V}}(\mathbf{x},t; \mathbf{w_F}) = \frac{1}{M} \sum_{i=0}^{M} \mathcal{A}_i(\mathbf{x}) \sin(2\pi\omega_i t) + \mathcal{B}_i(\mathbf{x}) \cos(2\pi\omega_i t)
\end{dmath}

Here, $\omega_i \sim \mathcal{N}(0,F_{max})$ is a random variable sampled from a normal distribution with standard deviation $F_{max}$. 
In practice, the maximum acceleration of natural objects will have physical constraints. 
This allows us to bandlimit the Fourier representation and parameterize the network with $F_{max}$.
We determined $F_{max}$ empirically with the goal of encouraging VelocityNet to represent more physically-relevant displacement fields.

EncoderNet and VelocityNet values are summed to estimate the SDF value of an object at every spatiotemporal position.  

\textbf{Implementation}
SIRENs \cite{sitzmann2020implicit}, an efficient framework capable of capturing high frequency information, were used to implement both EncoderNet and VelocityNet. 
As a result, the overall representation is given by
\begin{equation}
    \mathbf{\hat{g}}(\mathbf{x},t;\mathbf{w}) = \mathbf{\hat{g}_V}(\mathbf{x},t;\mathbf{w_F}) + \mathbf{\hat{g}_E}(\mathbf{x};\mathbf{w_E})
\end{equation}
where $\mathbf{w}$ is the union of EncoderNet weights $\mathbf{w_E}$ and VelocityNet weights $ \mathbf{w_F}$. i.e. $\mathbf{w}= \mathbf{w_E} \cup \mathbf{w_F}$.

\vspace{-10pt}
\subsection{Renderer}
In CT imaging, the acquired sinogram $r(l,\theta)$ represents the attenuation accumulated by x-rays traversing from the source to a specific detector position $l$ at a gantry position $\theta$. 

As described above, the gantry rotates over time $t$ such that, if an object has a spatialtemporal attenuation intensity $I(\mathbf{x},t)$, then the resultant sinogram value at detector position $l$ is given by:
\begin{equation}
    r(l,\theta) = \int_{u} I(\mathbf{x},t) \Gamma_{\theta}(t)du
\end{equation}
where $u$ is the path the ray traverses through the scene and $\Gamma_{\theta}(t) \in SO_2$ is the time-varying rotation matrix which describes the gantry rotation by an angle $\theta$ about the center of the scene.

In practice, image reconstruction is desired on an $N$-dimensional grid of size $d$. 
Therefore, the neural implicit SDF estimate can be discretized to obtain $\mathbf{\hat{f}_{\square}}(\mathbf{x}, t, k, \hat{g})$.
Occupancy of this discretized representation can be computed via $\zeta(x)$, defined as:
\begin{equation}
    \zeta(x) = min(1,max(0,\mu*(\sigma(x)-0.5)))
\end{equation}
where $\mu$ is a scaling factor that controls the sharpness of the boundary and $\sigma$ refers to the sigmoid function. 

Then, using the attenuation intensity of the object ${a}(k)$, we can estimate of the spatiotemporal intensity map $\hat{I}(\mathbf{x},t)$ as 
\begin{equation}
    \hat{I}(\mathbf{x},t)={a}(k)\zeta(\hat{f}_{\square}(\mathbf{x}, t, k, \hat{g}))
\end{equation}

The sinogram represents a discrete set of spatially- and angularly-integrated measurements in terms of both $l$ and $\theta$. 
Therefore, we performed both the forward rendering/projection operations at sub-pixel resolution to remove aliasing artifacts.
We did so by upsampling the derived occupancy grid $\mathbf{\hat{f}}(\mathbf{x},t;\mathbf{w})$, $k_{samp}$ times using bilinear interpolation.
Then, after calculating the projection using the above integral, $r(l,\theta)$ was downsampled via average pooling. 

The steps outlined above result in a differentiable rendering operator $R(\cdot;\theta) : \mathbb{R}^{d^N} \times \mathbb{R} \rightarrow \mathbb{R}^{d^{N-1}}$ which can map an SDF representation $\hat{g}(\mathbf{x},t;\mathbf{w})$ to the projection domain $\hat{r}(l,\theta)$ via appropriate discretization.

% \vspace{-5pt}
\section{How to reconstruct using NeuralCT}
\begin{figure*}[!htp]
\centering
\centerline{\includegraphics[width=6.5in]{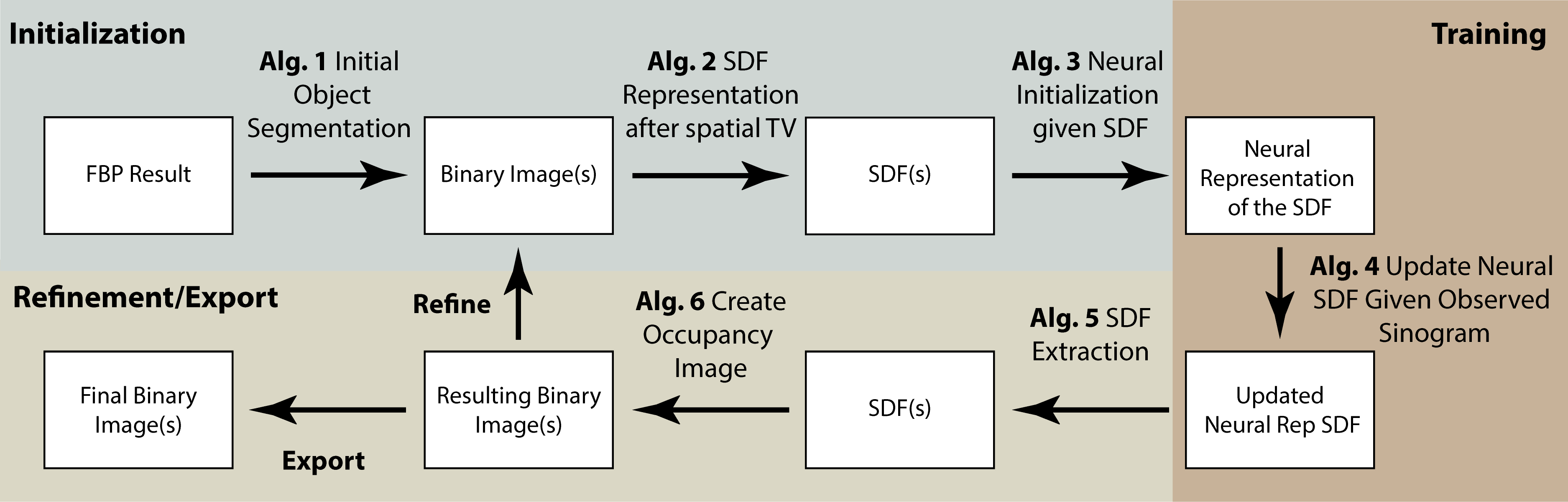}}
\caption{\textbf{Neural CT algorithm overview.} NeuralCT represents the spatiotemporal attenuation map of a moving object both explicitly in the image domain as well as implictly in the neural domain and performs optimization in both of these domains across three stages:  Initialization, Training, Refinement/Export. 
Initialization obtains well-behaved SDF representations based on filtered back projection (FBP) reconstructed images using an intensity-based segmentation, encodes the resulting segmentation as signed distance functions, and performs the explicit-to-implicit representation conversion.
Training aims to update neural SDF representation to match the sinogram data. Refinement/Export includes the implicit-to-explicit conversion and creation of occupancy images. 
NeuralCT results improved when it was repeated using the results of the first prediction. }
\label{fig_schematic}
\vspace{-10pt}
\end{figure*}

The design of NeuralCT leverages the fact that the spatiotemporal attenuation map of a moving object can be represented both explicitly in the image domain as well as implictly in the neural domain. 
That is, as weights of a neural network. 
Thus, NeuralCT performs optimization in both of these domains across three stages:  Initialization, Training, Refinement/Export as outlined in Figure \ref{fig_schematic}. 

Briefly, Initialization is intended to obtain well behaved gradients (that satisfy eikonal constraint \cite{sitzmann2020implicit}) instead of using a random initialization for the neural SDF representation. 
To do so, \textbf{Algorithm 1} identifies objects of interest in an image reconstructed via filtered back projection (FBP) using a intensity-based segmentation.
\textbf{Algorithms 2} encodes the resulting segmentation (binary background-foreground) image as signed distance functions and \textbf{Algorithm 3} performs the explicit-to-implicit representation conversion.

The Training portion begins with \textbf{Algorithm 4} which is responsible for updating (a.k.a. training) the neural SDF representation to match the sinogram data. 
Subsequently, \textbf{Algorithm 5} performs the implicit-to-explicit conversion and \textbf{Algorithm 6} creates the occupancy image for the discretized SDF map.

We observed improved results when NeuralCT was re-initialized using the result of it's first prediction. Therefore, Refinement/Export consists of applying \textbf{Algorithms 2 - 6} on the initial result of NeuralCT.

\vspace{-10pt}
\subsection{Initialization}
The SDF of $K$ foreground objects of interest is initialized using the filtered back projection (FBP) reconstruction images $I_{FBP}(\mathbf{x},t)$.
Here, the number of objects of interest $K$ is defined apriori.
Defining both a background class as well as several additional classes $\kappa$ such that $K'=K+\kappa+1$ improved initialization. 

\textbf{Algorithm 1} performs intensity-based segmentation using a Gaussian Mixture Model \cite{reynolds2009gaussian} $GMM(I_{FBP}(\mathbf{x},t),k')$ to create a segmentation images $G(\mathbf{x},t)$.
This results in binary classification images $C(\mathbf{x},t,k')$.
From these binary images, the mass $\mathcal{M}(k')$ of each foreground class was calculated.
After removal of the background $c_{background}$, defined as the class with the largest mass, the binary images $C(\mathbf{x},t,k)$ for the $K$ largest classes were kept.

\setlength{\textfloatsep}{0pt}
\begin{algorithm}[!tp]
    \SetKwInOut{Input}{Input}
    \SetKwInOut{Output}{Output}

    \Input{FBP images $I_{FBP}(\mathbf{x},t)\forall t\in T$, \\
     number of objects $K$, \\
     buffer classes $\kappa$}
    \Output{Binary classification images $C(\mathbf{x},t,k)\forall t\in T$}
    
    {
        Randomly, select $\tau \subset T$ \\
        $Segmentor \leftarrow$ GMM$(I_{FBP}(\mathbf{x},t),k') \ t \in \tau$ \\
        $G(\mathbf{x},t,k') \leftarrow Segmentor$.fit$(I_{FBP}(\mathbf{x},t)\forall t\in T)$ \\
        
        $\mathcal{M} \leftarrow \left\{ \cdots, \sum_{x\in\Omega}G(\mathbf{x},t,c_i),\cdots \right\}$
        
        $c_{background} \leftarrow \underset{c}{\mathrm{argmax}} \ \mathcal{M}$ \\
        
        $\mathcal{M} \leftarrow \mathcal{M}$.delete($c_{background}$) \\
        $\left\{ c_i\right\} \leftarrow $ TopKIndices($\mathcal{M}$) \\
        
        $C(\mathbf{x},t,k) \leftarrow G(\mathbf{x},t,\left\{ c_i\right\})$ \\
    }
    {
        return $C(\mathbf{x},t,k)$\;
    }
    \caption{Object Segmentation}
    \label{alg1_object_seg}
\end{algorithm}

\begin{algorithm}[!tp]
    \SetKwInOut{Input}{Input}
    \SetKwInOut{Output}{Output}

    \Input{Binary class images $C(\mathbf{x},t,k)\forall t\in T$}
    \Output{SDF images $\hat{f}(\mathbf{x},t,k)\forall t\in T$}
    
    {
        $\Tilde{B}(\mathbf{x},t,k) \leftarrow $ TotalVariationMinimizer($C(\mathbf{x},t,k)$) \\
        
        $\Tilde{D}(\mathbf{x},t,k) \leftarrow $ DistanceTransform($\Tilde{B}(\mathbf{x},t,k)$) \\
        
         $\hat{f}(\mathbf{x},t,k) \leftarrow $ TotalVariationMinimizer($\Tilde{D}(\mathbf{x},t,k)$) \\
    }
    
    {
        return $\hat{f}(\mathbf{x},t,k)$\;
    }
    \caption{Convert Binary images to SDF}
    \label{alg2_conv2binary}
\end{algorithm}

Binary class images $C(\mathbf{x},t,k)$ representing pixels of each label across time were then converted into SDF images $\hat{f}(\mathbf{x},t,k)$ using \textbf{Algorithm 2}.
Our approach uses total variation minimization \cite{osher2005iterative} to smooth the images and then performs the distance transform over the entire domain to obtain SDF images $\hat{f}(\mathbf{x},t,k)$. 

To create neural implicit representations from the explicit SDF images $\hat{f}(\mathbf{x},t,k)$, we performed optimization as described in \textbf{Algorithm 3}. 
A randomly initialized neural network $\mathbf{g}(\mathbf{x},t;\mathbf{w})$ is optimized into a network $\mathbf{\hat{g}_{o}}(\mathbf{x},t;\mathbf{w})$ which best approximates the explicit SDF image $\mathbf{\hat{g}}(\mathbf{x},t;\mathbf{w}) \approx \hat{f}(\mathbf{x},t,k)$. 
The optimization is directly supervised and aims to minimize the SDF differences $\mathcal{L}_{SDF}$ and satisfy the Eikonal constraint $\mathcal{L}_{Eikonal}$.
The optimization ends when the maximum number of iterations $maxIterations$ have been reached. 

\begin{algorithm}[!tp]
    \SetKwInOut{Input}{Input}
    \SetKwInOut{Output}{Output}

    \Input{Neural SDF $\mathbf{g}(\mathbf{x},t;\mathbf{w})$, \\
    SDF images $\hat{f}(\mathbf{x},t,k)$}
    
    \Output{Initialized neural SDF $\mathbf{\hat{g}_{o}}(\mathbf{x},t;\mathbf{w}) \forall t\in T$}
    
    $\mathbf{\hat{g}}(\mathbf{x},t;\mathbf{w})=\mathbf{g}(\mathbf{x},t;\ $rand$(\mathbf{w}))$ \\
    
    \For{$iteration \ < \ maxIterations$ }
    {
        Sample $t \sim T$ \\
        
        $\mathcal{L}_{SDF} \leftarrow \frac{1}{|\Omega|}\sum_{x\in\Omega}\left |\mathbf{\hat{g}}(\mathbf{x},t;\mathbf{w})-\hat{f}(\mathbf{x},t) \right |$ \\
        
        $\mathcal{L}_{Eikonal} \leftarrow \frac{1}{|\Omega|}\sum_{x\in\Omega}\left |||\nabla_x\mathbf{\hat{g}}(\mathbf{x},t;\mathbf{w})||_2 - 1 \right |$ \\
        
        $\mathcal{L} \leftarrow \mathcal{L}_{SDF} + \lambda \mathcal{L}_{Eikonal}$ \\

        $\mathbf{w} \leftarrow \mathbf{w} - \alpha\nabla \mathcal{L}$ \\
        
        % \If{$\mathcal{L} < minLoss$}
        %     {break}
    }
    $\mathbf{\hat{g}_{o}}(\mathbf{x},t;\mathbf{w})\leftarrow \mathbf{\hat{g}}(\mathbf{x},t;\mathbf{w})$ \\
    {
        return $\mathbf{\hat{g}_{o}}(\mathbf{x},t;\mathbf{w})$; \\
    }
    \caption{Initializing Neural SDF Representation}
    \label{alg3_init_neuralSDF}
\end{algorithm}

\begin{algorithm}[!h]
    \SetKwInOut{Input}{Input}
    \SetKwInOut{Output}{Output}

    % \underline{function Euclid} $(a,b)$\;
    \Input{Initialized SDF  $\mathbf{\hat{g}_{o}}(\mathbf{x},t;\mathbf{w})\forall t\in T$, acquired sinogram $r(\mathbf{l},\bm{\theta})$}
    
    \Output{Optimized SDF $\mathbf{\hat{g}}(\mathbf{x},t;\mathbf{w}) \forall t\in T$}
    
    \For{$iteration \ < \ maxIterations$ }
    {
        Sample $t \sim T$ \;
        
        Compute projection $\hat{r}(\mathbf{l},\bm{\theta}) \leftarrow R(\hat{I}(\mathbf{x},t),\theta(t))$ \;
        
        $\mathcal{L}_{Sinogram} \leftarrow \frac{1}{|\Omega|}\sum_{x\in\Omega}\left | r(\mathbf{l},\bm{\theta})-\hat{r}(\mathbf{l},\bm{\theta}) \right |$ \\
        
        $\mathcal{L}_{Eikonal} \leftarrow \frac{1}{|\Omega|}\sum_{x\in\Omega}\left |||\nabla_x\mathbf{\hat{g}_{o}}(\mathbf{x},t;\mathbf{w})||_2 - 1 \right |$ \\
        
        $\mathcal{L}_{TVS} \leftarrow \frac{1}{|\Omega|}\sum_{x\in\Omega}||\nabla_x\mathbf{\hat{g}_{o}}(\mathbf{x},t;\mathbf{w})||_1$ \\
        
        $\mathcal{L}_{TVT} \leftarrow \frac{1}{|\Omega|}\sum_{x\in\Omega}||\nabla_t\mathbf{\hat{g}_{o}}(\mathbf{x},t;\mathbf{w})||_1$ \\
        
        $\mathcal{L} \leftarrow \mathcal{L}_{Sinogram} + \lambda_1 \mathcal{L}_{Eikonal} + \lambda_2 \mathcal{L}_{TVS} + \lambda_3 \mathcal{L}_{TVT}$ \\
        
        $\mathbf{w} \leftarrow \mathbf{w} - \alpha\nabla \mathcal{L}$ \\
            
    \If{$\mathcal{L}_{Sinogram} < minLoss$}
        {break}
    }
    
    $\mathbf{\hat{g}}(\mathbf{x},t;\mathbf{w}) \leftarrow \mathbf{\hat{g}_{0}}(\mathbf{x},t;\mathbf{w})$ \;
    {
        return $\mathbf{\hat{g}}(\mathbf{x},t;\mathbf{w})$\;
    }
    \caption{Optimizing Neural SDF Representation}
    \label{alg4_opt_neuralSDF}
\end{algorithm}

\vspace{-10pt}
\subsection{Training}
At this point, the SDF representation $\mathbf{\hat{g}_{o}}(\mathbf{x},t;\mathbf{w})$ includes motion artifacts present in $I_{FBP}$. 
\textbf{Algorithm 4} optimizes the neural network SDF representation $\mathbf{\hat{g}}(\mathbf{x},t;\mathbf{w})$ to best explain the acquired sinogram $r(\mathbf{l},\bm{\theta})$ with the minimum total variation in space $||\nabla_x\mathbf{\hat{g}}(\mathbf{x},t;\mathbf{w})||_1$ and time $||\nabla_t\mathbf{\hat{g}}(\mathbf{x},t;\mathbf{w})||_1$. 

The current neural implicit SDF prediction $\mathbf{\hat{g}}(\mathbf{x},t;\mathbf{w})$ can be converted to a spatiotemporal intensity map $\hat{I}(\mathbf{x},t)$ and projected to the sinogram domain via previously described renderer (Section \ref{sec:NeuralCT}) $R(I(\mathbf{x},t),\bm{\theta})$.
This results in a sinogram estimate $\hat{r}(\mathbf{l},\bm{\theta})$ used to calculate the sinogram loss $\mathcal{L}_{Sinogram}$ —  the difference between the current estimate and the acquired sinogram $r(\mathbf{l},\bm{\theta})$. 
This loss is combined with the Eikonal constraint $\mathcal{L}_{Eikonal}$ and spatial $\mathcal{L}_{TVS}$ and temporal TV $\mathcal{L}_{TVT}$ loss terms to improve optimization. 
The optimization was performed until the projection loss $\mathcal{L}_{Sinogram}$ decreased below a certain threshold $minLoss$ or maximum number of iterations $maxIterations$ was reached.

\vspace{-10pt}
\subsection{Refinement/Exporting}

\begin{figure*}[!htp]
\includegraphics[width=\linewidth]{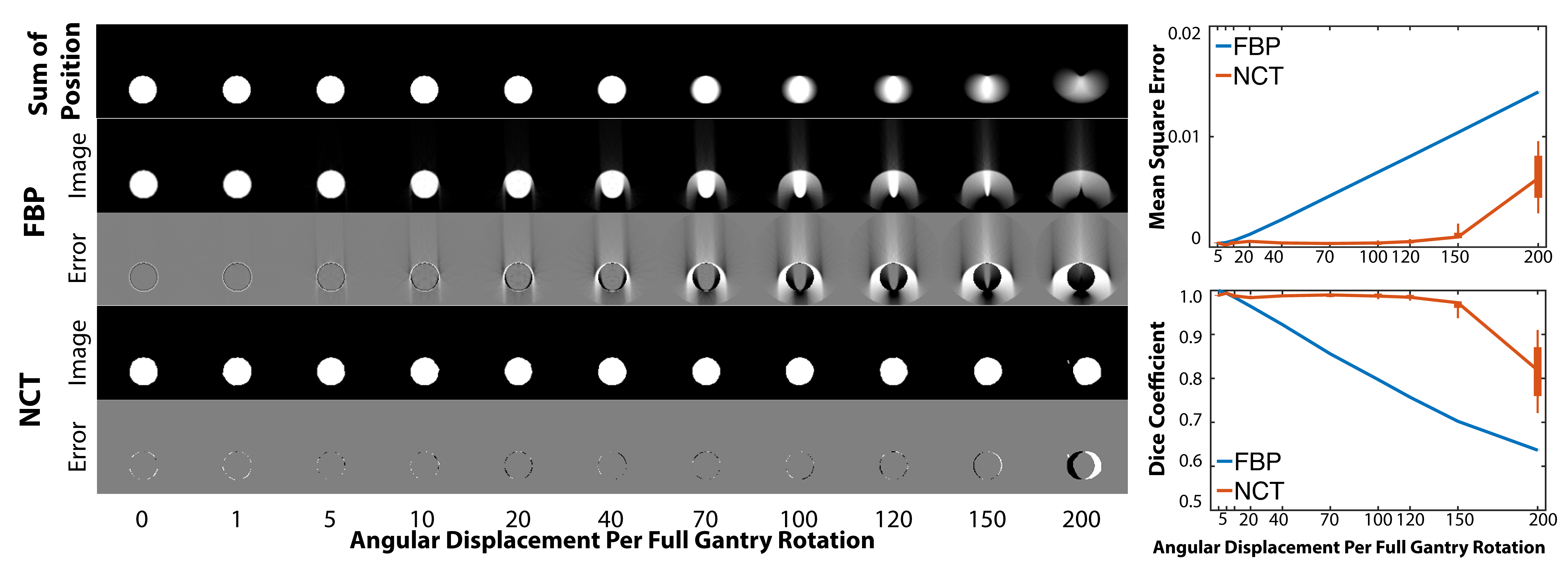}
\caption{
\textbf{NeuralCT (NCT) accurately depicts the position of a circle, despite high angular motion during acquisition.} 
FBP images are degraded with increasing motion during data acquisition. 
NCT improves delineation of the translating vessel, both visually and via Mean Squared Error and Dice Coefficient metrics.
Error bars represent the standard deviation observed from 5  NCT results using different random initialization of the networks.
All images are reconstructions using 360-degrees of projections. 
Displacement was imparted as an angular translation (in degrees) as indicated by the legend.}
\label{fig_dv_gantry0}
\vspace{-10pt}
\end{figure*}

To generate spatiotemporal intensity images from the neural SDF representation, we first convert the neural SDF into a set of explicit SDF images $\hat{f}_{\square}(\mathbf{x},t,k)$.
This is achieved by sampling the neural SDF $\mathbf{\hat{g}}(\mathbf{x},t;\mathbf{w})$ over a $n$-dimensional grid at a desired spatial resolution (\textbf{Algorithm 5}). 
The resulting SDF image is then binarized $\hat{B}(\mathbf{x},t,k) = \zeta(\hat{f}_{\square}(\mathbf{x},t,k))$ (\textbf{Algorithm 6}). 
Binarized images $\hat{B}(x,t,k;w)$ are then used for Refinement, a second pass through the algorithm (starting with \textbf{Algorithm 2}).
After Refinement, images were exported and analyzed.

% \vspace{-5pt}
\section{Experiments}
\label{sec:experiments}

\begin{figure}[!ht]
\centering
\includegraphics[width=\columnwidth]{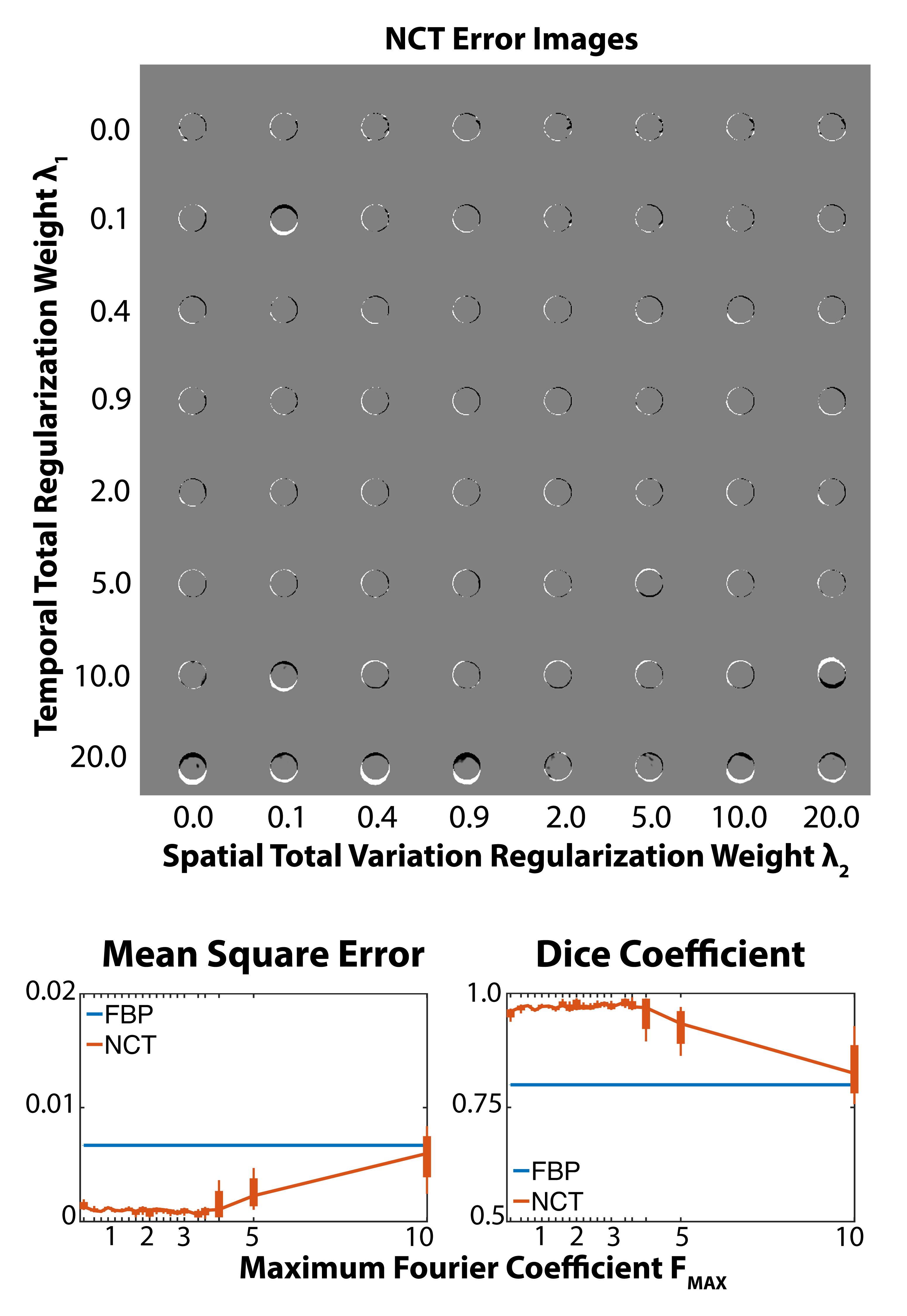}
\caption{\textbf{NeuralCT (NCT) reconstruction of an object with angular displacement = $100^\circ$ with varying spatial and temporal regularization (top) and $F_{max}$ (bottom).} NCT yields accurate image reconstruction over a range of spatial TV regularization. However, NCT reconstruct can become inaccurate if temporal regularization is too highly penalized. This corresponds to making the solution more stationary. NCT results are also robust for a wide-range of $F_{max}$. At $F_{max}>10$, the increased parameterization can lead to overfitting but robust results can be achieved with lower $F_{max}$.}
\label{fig_spat_temp_reg}
% \vspace{-15pt}
\end{figure}

\begin{figure*}[!ht]
\centering
\includegraphics[width=\linewidth]{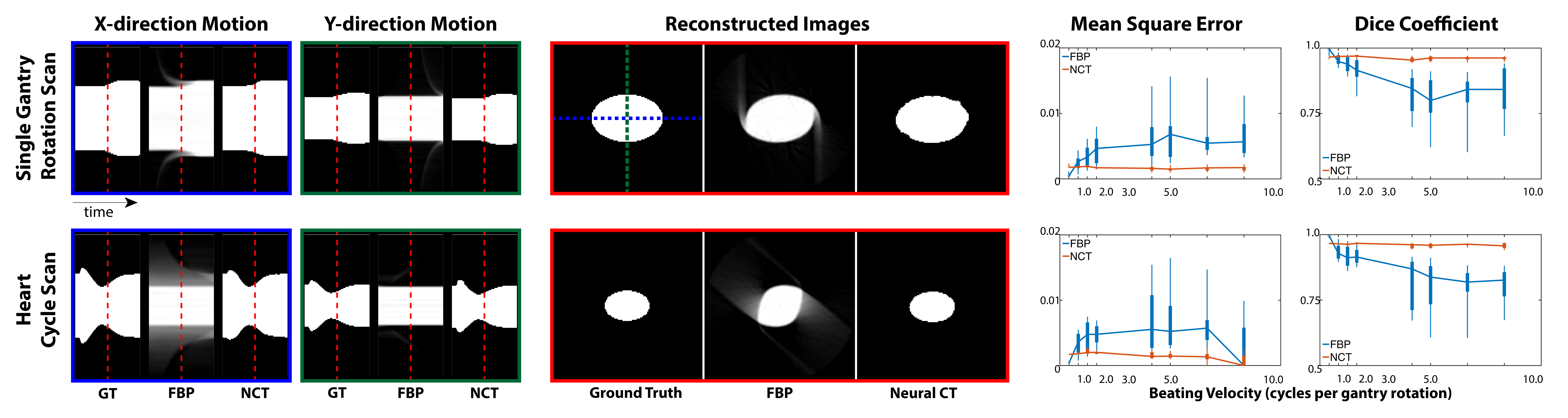}
\caption{\textbf{NeuralCT (NCT) improves single (top) and multiple (bottom) gantry rotation imaging.} 
\textbf{Top:} NCT improves image quality, relative to FBP for both cases. The motion of the X (blue) and Y (green) axes is shown to illustrate how FBP blurring is improved by NCT. 
Images reconstructed at the middle of the data acquisition (red line) are shown in the middle panel.
\textbf{Bottom:} NCT leverages the additional temporal information available when multiple gantry rotations are acquired and improves image quality, over FBP, without modification.}
\label{fig_beating_heart}
\vspace{-10pt}
\end{figure*}

%\vspace{-10pt}
\subsection{Implementation Details}
The image resolution used during training was set to $n=128$ with upsampling factor $k_{samp}=2$. 
For optimal performance, we used $F_{max}=3.0$ and $\mu=50$. 
In \textbf{Algorithm 1}, we sampled $\tau$ such that $|\tau| = 0.02*|T|$ and set $\kappa=3$. 
For \textbf{Algorithms 3 \& 4} the learning rate was $\alpha=1\times 10^{-5}$ and decayed by a factor of $0.95$ every $200$ steps. 
The optimization procedures was run with $maxIterations=5000$, $minLoss=0.08$, and $\lambda=0.1$. For best results, we trained with minibatch $|t| = 20$. 
As described below, we observed optimal results with $\lambda_1=0.1, \lambda_2=0.5$, and $\lambda_3=0.5$. 
All experiments were performed using a 2D parallel beam CT geometry of a single foreground class (i.e. $K=1$). Optimizing NeuralCT takes $\sim 30$ mins on a single NVIDIA 1080Ti GPU.

\vspace{-10pt}
\subsection{Evaluation of NeuralCT Parameters}
First, we simulated a simple cause of motion corruption: imaging a bright circle undergoing translation. 
This was intended to mimic motion of a coronary artery \cite{contijoch_medphys_2017}.
We used this toy problem to explore the impact of $\lambda_1$, $\lambda_2$, and $F_{max}$ on the ability of NeuralCT to faithfully reconstruct a moving object.
$\lambda_1$ and $\lambda_2$ are the weight of spatial and temporal total variation in the cost function used to supervise learning and $F_{max}$ is the maximum Fourier coefficient available for representation of the temporal changes.

The impact of these parameters on NeuralCT was evaluated via parameter sweeps.
First, $\lambda_1$ and $\lambda_2$ were varied (0, 0.05, 0.1, 0.15, 0.3, 0.5, 0.9, 1.5, 3.0, 5.0, 10.0) with $F_{max}$ set to 3.0.
Then, $F_{max}$ was varied from 0 to 10.0 with $\lambda_1$=0.5 and $\lambda_2$=0.5, 

We evaluated a range of angular displacements per gantry rotation (0, 1, 5, 10, 20, 40, 70, 100, 120, 150, and 200 degrees per gantry rotation).
NeuralCT results were compared to both the ground-truth vessel image as well as the FBP result using the metrics described below.

\vspace{-10pt}
\subsection{Imaging of Nonrigid, Heart-like Motion}
To evaluate the utility of NeuralCT during non-rigid motion, we modeled heart-like motion using an ellipse with time-varying axes.
In addition to static imaging, cine CT imaging can be used to evaluate heart size and dynamics. 
Therefore, we evaluated two imaging approaches: 1) Single-gantry rotation imagin where a sinogram spanning 360 degrees is acquired during which the ellipse diameter changes.  
2) Full heart cycle imaging where multiple (typically 4-5) gantry rotations were obtained spanning the entire period of the heart-like motion.
Of note, NeuralCT readily incorporate multiple rotation data into the reconstruction framework without modification.
In this scenario, NeuralCT results were compared to FBP reconstructions centered at the same temporal position.

\vspace{-10pt}
\subsection{Imaging of Complex Deformations}
Lastly, we demonstrate the ability of NeuralCT to reconstruct scenes with complex topological change.
To do so, we created a complex scene where the letter ``A" transforms into ``B", then into ``C", then back to ``B' and ``A".
Without changes to any NeuralCT parameters (spatial/temporal weighting and Fourier coefficients), NeuralCT results were compared to FBP imaging. 

\vspace{-10pt}
\subsection{Metrics to Evaluate Image Quality}
NeuralCT was compared to conventional FBP images  using mean-square error (MSE) and the foreground's Dice coefficient where FBP images were thresholded using half of the foreground intensity.
We did not compare NeuralCT to motion correction methods.
This was motivated by the fact that the improvement obtained via correction depends on the suitability of the motion model in the correction scheme to the motion present in the acquired data.
Given that current correction methods have been designed for clinical scenarios, our motion scenarios are not expected to represent realistic use cases.
For NeuralCT, we report the average of 5 independent optimizations, each initialized with a different random seed.

\begin{figure}[!ht]
\centerline{\includegraphics[width=\columnwidth]{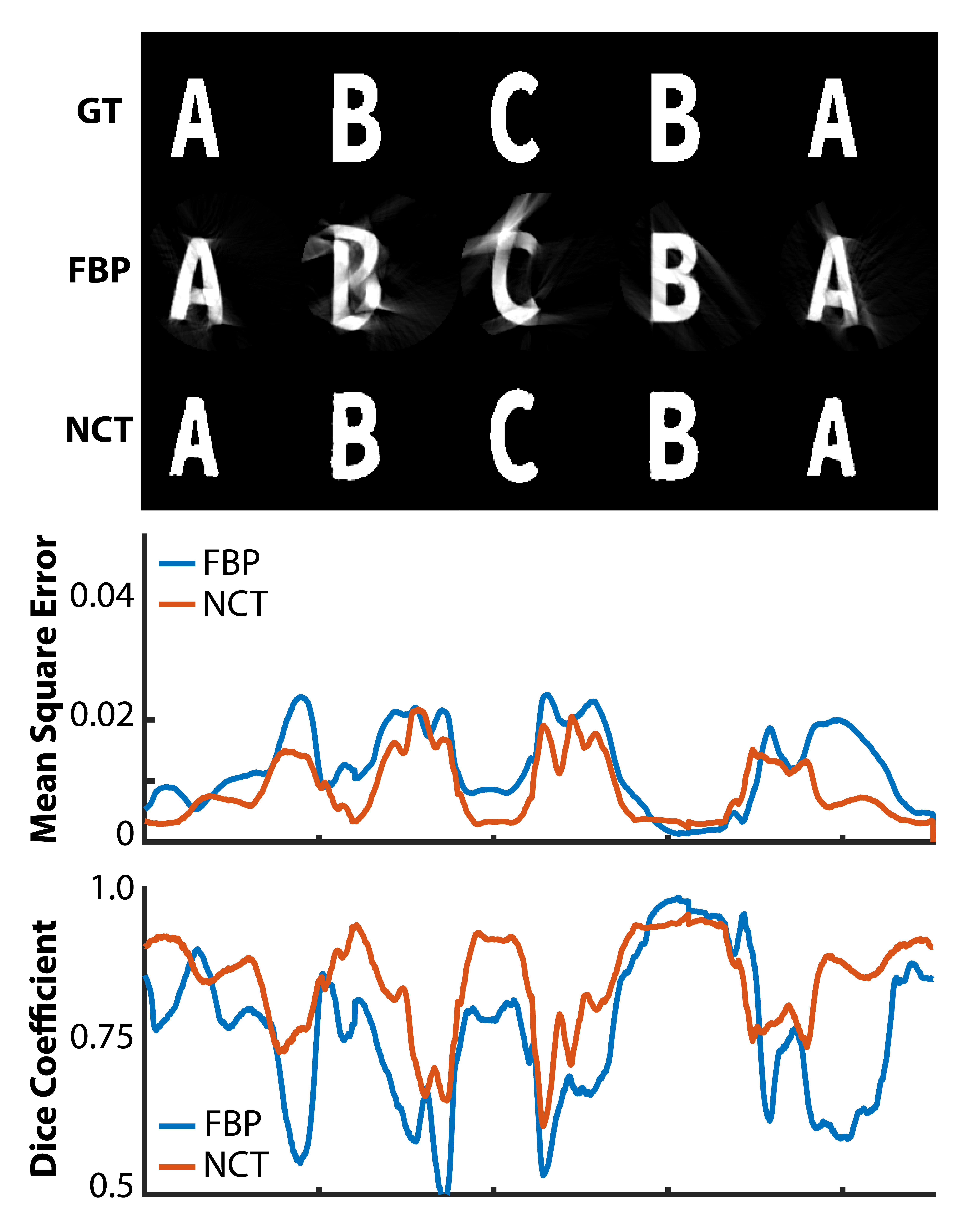}}
\caption{\textbf{Evaluation of NeuralCT (NCT) when imaging during a complex change in topology.} Without modification of the framework or tuning of parameters, NCT improved imaging of a complex scene. Further, the approach did so without estimation of motion or a prior information.}
\label{fig_letter_warp}
%\vspace{-20pt}
\end{figure}

% \vspace{-5pt}
\section{Results}
\label{sec:results}
\subsection{Moving Coronary Vessel Imaging}
As shown by the images and metrics in Figure \ref{fig_dv_gantry0}, NeuralCT accurately reconstructed images of a small circle while FBP images showed significant artifacts ($\lambda_1=0.5$, $\lambda_2=0.5$, and $F_{max}=3.0$). 
Low MSE and high ($>0.9$) Dice was maintained for angular displacements of up to 150$^\circ$ during data acquisition.
Figure \ref{fig_dv_gantry0} also illustrates how the NeuralCT reconstruction is impacted by displacements \textgreater$150^\circ$. 
While the shape of the circle is fairly preserved,  increased MSE and decreased Dice occur due to reconstruction of the circle at the incorrect temporal position.

\vspace{-5pt}
\subsection{Evaluation of NeuralCT Parameters}
NeuralCT reconstruction of a circle with translation = $100^\circ$ per gantry rotation was used to evaluated the robustness of NCT to changes in $\lambda_1$, $\lambda_2$, and $F_{max}$.
Figure \ref{fig_spat_temp_reg} (top) illustrates the minor impact of changing the strength of spatial ($\lambda_1$) and temporal ($\lambda_2$) regularization while maintaining $F_{max}$ = 3.0.
Accurate reconstruction was achieved over a wide range of regularization strength.
Figure \ref{fig_spat_temp_reg} (bottom) illustrates the performance of NeuralCT with $\lambda_1=0.5$ and $\lambda_2=0.5$ when varying $F_{max}$.
Decreasing $F_{max}$ limits the temporal evolution that can be represented by the neural implicit approach. 
However, we observed robust NeuralCT reconstruction performance even at low $F_{max}$ values.
Reconstruction accuracy decreased at high $F_{max}$.
This suggests that introduction of high frequency coefficients can result in overfitting.

\vspace{-5pt}
\subsection{Imaging Nonrigid, Heart-like Motion with NeuralCT}
Without modification of the parameters identified above ($\lambda_1=0.5$, $\lambda_2=0.5$, and $F_{max}=3.0$), NCT successfully reconstructed images of the heart-like motion for both single gantry and full heart cycle imaging.
Relative to FBP, NeuralCT improved imaging metrics when imaging an ellipse with changing dimensions.
Figure \ref{fig_beating_heart} illustrates these differences.
The change in axes dimension is shown by the left column (temporal reformats along the x- and y-direction). 
Metrics of image quality (MSE and Dice) are shown on the right. 
For full heart cycle imaging with beating velocity = 3.0, NCT decreased MSE (FBP: median 0.005, IQR 0.003-0.009, NCT: median 0.001, IQR 0.001-0.002).
Further, NCT increased the percentage of frames with MSE $<0.005$ from 46\% (FBP) to 100\%.
NCT also improved Dice (FBP: median 0.84, IQR 0.73-0.88 to NCT: median 0.96, IQR 0.95-0.97). 
Supplemental Movie 1 illustrates the ground truth motion and improvement of NeuralCT reconstruction, relative to FBP.

\vspace{-10pt}
\subsection{Imaging Complex Topology Changes with NeuralCT} 
Without modification of the NeuralCT framework or parameters ($\lambda_1=0.5$, $\lambda_2=0.5$, and $F_{max}=3.0$), NCT successfully reconstructed data acquired during a complex letter warping scene.
Figure \ref{fig_letter_warp} shows the stationary periods of the scene with both the groundtruth (top) as well as FBP (middle) and NeuralCT (bottom) reconstructions.
NeuralCT significantly reduced the severity of artifacts observed with FBP.
The plot of MSE and Dice scores as a function of time further illustrate the improvement.
NCT decreased error as measured via MSE (FBP: median 0.011, IR 0.008-0.019 to NCT: median 0.007, IQR 0.004-0.013). 
Further, NCT increased the percentage of the frames with MSE$<0.005$ from 15.2\% to 34\%.
DICE scores also improved with NCT (FBP: median 0.78, IQR 0.66-0.85, NCT: median 0.86, (IQR 0.80-0.91). 
The percentage of frames with Dice $>0.85$ increased from 25.8\% for FBP to 58.8\% with NCT.
Supplemental Movie 2 illustrates the ground truth and FBP and NeuralCT reconstruction of the complex scene.

% \vspace{-5pt}
\section{Discussion}
\label{sec:discussion}
NeuralCT successfully combines implicit representation of an object by a signed distance function (SDF) with differentiable rendering to enable time-resolved imaging free from motion artifacts despite data acquisition occuring during motion.
NeuralCT takes advantage of several important features of SDFs -- namely, that they represent movement of a boundary as a spatially and temporally smooth evolution.
NeuralCT represents the scene as SDFs which evolve over time using an implicit representation; in this case, a neural network.
Differentiable rendering is used to improve the estimate of the scene by comparing the observed CT data with the SDF-based representation in the sinogram domain.
The framework also enables additional regularization such as penalizing deviations from the Eikonal constraint and minimizing spatial and temporal variations.
We demonstrated the utility of NeuralCT in three different imaging scenarios without changes to the architecture. 
Specifically, NeuralCT was readily applied to objects with different motions as well as data spanning one or more than one gantry rotation.
These cases highlight how NeuralCT can be used to accurately reconstruct objects undergoing 1) translation, 2) heartbeat-like affine changes in diameter, and 3) complex topological changes (warping of letters).
This flexibility was facilitated by the fact that NeuralCT does not utilize an explicit motion model.

In our evaluation of NeuralCT, we observed that accurate reconstruction were possible even with a fairly limited set of Fourier coefficients.
This is likely due to the beneficial fact that SDFs used to parameterize the reconstruction problem evolve smoothly over space and time.
We also observed reconstruction performance decreased if $F_{max}$ increased. 
This highlights the benefit of limiting the complexity of the solution via  $F_{max}$. 
An added advantage is that this hyperparameter has a simple physical interpretation, the bandwidth of the motion, which could facilitate its selection when applying NeuralCT to novel applications.
Further, in our examples, a single foreground object was imaged with an empty background. 
However, in practice, the background class can be used to represent the position and intensity of static objects.

We illustrate NeuralCT results without modification of parameters $\lambda_1,\lambda_2$, and $F_{max}$.
However, the optimal choice of these parameters is expected to vary depending on the specific imaging scenarios.
Further, the rotation speed of CT systems can depend on both the design of the system and the clinical protocol.
Evaluating the optimal combination of parameters is left for future work where specific scenarios are evaluated.

In this manuscript, we illustrate how NeuralCT can seamlessly reconstruct data acquired using two common CT acquisition approaches (single and multiple gantry rotation imaging).
Specifically, NeuralCT reconstructed imaging data spanning multiple gantry rotations simultaneously without explicit labeling of timepoints or the need for specific, complementary images.
This is significantly different than approaches such as FBP which reconstruct each frame independently.
As a result, in addition to improving single gantry reconstruction, NeuralCT also has the potential to improve reconstruction quality of each frame acquired during a multiple gantry acquisition by leveraging additional temporal information.
While acquiring additional temporal information (beyond the typical half- or single gantry rotation) increases the dose delivered to the patient, it could enable NeuralCT to resolve the dynamics of objects that have been significantly hampered by imaging artifacts.

We believe our findings are noteworthy and novel but our study had several limitations.
First, we quantified NeuralCT image quality relative to the groundtruth and conventional FBP reconstruction instead of direct comparison to previously-reported motion-correction approaches.
We chose not to compare NeuralCT to techniques specially crafted for certain imaging scenarios such as translation of a coronary artery.
We did so because we expect performance of motion correction approaches to depend significantly on the specifics of the application.
For example, we expect motion-correction approaches to accurately reconstruct scenes when the object motion agrees with the method's motion model. 
However, limited performance may occur if the object motion is different from the motion model.
While our approach does not require an a priori motion model, it is difficult to ensure that our examples adhere to the constraints incorporated into current approaches.
Comparison to current motion-correction algorithms is planned for future work in specific clinical scenarios.
Second, as indicated above, we demonstrated the use of NeuralCT using a single foreground class and imaging with a single-source 2D parallel beam geometry. 
This was done to simplify interpretability of our findings.
While dual-source systems, fan- or cone-beam geometry, and multiple object scenes will change the relationship between object motion and acquired CT data, we expect NeuralCT to improve image quality in these scenarios. 
However, these extensions are left for future work.

% \vspace{-5pt}
\section{Conclusions}
\label{sec:conclusions}
A novel reconstruction framework, NeuralCT, can be used to reconstruct CT imaging data acquired during object motion in a time-resolved manner free from motion artifacts. 
Our approach leverages a neural implicit scheme and does not require a prior motion models or explicit motion estimation. 
Representing moving boundaries using a signed distance metric and neural implicit framework enables `analysis-by-synthesis' to identify a solution consistent with the observed sinogram as well as spatial and temporal consistency constraints. 

% \vspace{-5pt}
\section*{Description of Supplemental Movies}
\textbf{Supplemental Movie 1: Visualization of beat-like motion from Figure \ref{fig_beating_heart}}. NeuralCT (right) improves image quality over filtered backprojection (left) when reconstructing an ellipse with time-varying dimensions. Center: ground truth.

\textbf{Supplemental Movie 2: Visualization of letter warping from Figure \ref{fig_letter_warp}}. NeuralCT (right) improves imaging of a complex topological change relative to filtered backprojection (left). Center: ground truth.

% \vspace{-5pt}
\section*{Acknowledgment}
We thank Zhennong Chen for comments and insightful discussion regarding the manuscript.

\bibliographystyle{IEEEtran}
% \vspace{-5pt}
\bibliography{References}

\end{document}